\documentclass[ reprint, superscriptaddress,
 amsmath,amssymb,
 aps,
 pra
]{revtex4-2}

\usepackage{graphicx}
\usepackage{overpic}
\usepackage{dcolumn}
\usepackage{bm}
\usepackage[hidelinks,colorlinks=true,linkcolor=blue,citecolor=blue]{hyperref}
\usepackage[mathlines]{lineno}
\usepackage[T1]{fontenc}
\usepackage{braket}
\usepackage{color}
\usepackage{float}

\begin{document}

\title{Analytical solutions of the Schr\"{o}dinger equation for two confined atoms with van der Waals interaction}

\author{Ruijie Du}
\email{2015201087@ruc.edu.cn}
\affiliation{Department of Physics, Renmin University of China, Beijing, 100872,
China}



%

\date{\today}

\begin{abstract}
We derive solutions of the Schr\"{o}dinger equation for the isotropic van der Waals interaction in a symmetric harmonic trap, with the recent approach [arXiv:2207.09377 (2022)] to handle the multi-scale long-range potential. Asymptotic behaviors of these solutions are then obtained for $r\rightarrow 0$ and $r\rightarrow \infty$. We further deduce the energy spectrum of the two-body relative motion and relate the spectrum to scattering lengths for $s$ wave and $p$ wave. These results can be used to research trapped atom-atom collisions and energy spectra.
\end{abstract}

\maketitle

\section{Introduction} 

Trapped atoms have been the fundamental ingredient of experiments in atomic and molecular physics \cite{Baly2000, Hut2006, Gar2016}. The relevant theoretical research \cite{Blume2012, Xiao2022} of two interacting atoms with confinement is thus a vital class of problems, which is the basis of interacting polyatomic systems. There are two competitive length scales in such systems: characteristic lengths of the interatomic interaction and the trap. So far, interacting atoms with confinement are mostly dealt with pseudopotential theories ({\it e.g.}, Huang-Yang pseudopotential \cite{Yang1957} and other generalizations \cite{Paul2002, Blume2004, ZI2009}). These theories are limited by the condition that the characteristic length of the interaction is much less than that of the trap, just as in most cases. For instance, the length scale of the van der Waals (vdW) interaction for widely-used alkali metal atoms is not more than the order of 100 nm \cite{Tao2013} while the typical size of magnetic or optical traps is of the order of 1 $\mu$m, which is obviously greater. Moreover, such theories are more suitable for $s$ wave.

However, the contrary situation ({\it i.e.}, the length scale of the interaction is comparable with or even greater than that of the trap) is obliged to be considered due to the development of sub-wavelength traps \cite{Yang2021} (including optical tweezers \cite{NKK2021}) and Rydberg atoms \cite{Lim2013, Jan2023}. The former are the central tool to precisely prepare and manipulate untracold neutral particles, and the latter contribute to quantum information and quantum simulation \cite{Shi2022, Wu2021}. Specifically, the size of sub-wavelength trap can be confined to the order of 10 nm \cite{Poul2009, Regal2012, Wang2018}, while the characteristic length for the vdW interaction between Rydberg atoms with high enough principal quantum numbers can vary from 1 $\mu$m to 100 $\mu$m \cite{Potv2012, Zhu2021}. In such a situation, pseudopotential theories fail to address the multi-scale physics.

Bo Gao {\it et al.} \cite{Bo2007} have made a successful attempt to solve the Schr\"{o}dinger equation (SE) for an isotropic vdW potential in a symmetric harmonic trap restricted by $a_h/\beta_6\gg 1$, where $a_h$ and $\beta_6$ are characteristic lengths of the harmonic trap and vdW potential respectively. Such a restriction can be released with our recent approach \cite{Du2022} to handle with the multi-scale long-range potential. 

In this paper, we solve the SE for two confined atoms in a symmetric harmonic trap with vdW interaction. A pair of linearly independent solutions to the radial SE is presented, as well as corresponding asymptotic behaviors in the limits $r\rightarrow 0$ and $r\rightarrow\infty$, with $r$ being the relative distance. According to the special solutions, we derive the energy spectrum of the two-body relative motion and the relation to the scattering length for $s$ wave and $p$ wave. These results can be used to research trapped atom-atom collisions and energy spectra.

The remainder of this paper is organized as follows. In Sec.~\ref{II} we give a pair of linearly independent solutions of radial SE and the asymptotic behaviors. The energy spectrum is derived in Sec.~\ref{III}. Other applications for our solutions and a brief summary are presented in Sec.~\ref{IV}. Some details of our derivations are illustrated in the appendixes.

\section{Solutions of the radial Schr\"Odinger equation} 
\label{II}

\begin{figure}[t]
\centering
\begin{overpic}[width=0.18\textwidth]{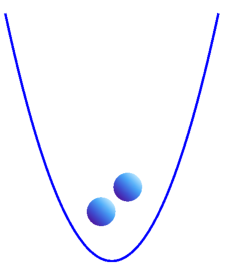}
\put(16,87){{\bf (a)}}
\end{overpic}\hspace{0.5cm}
\begin{overpic}[width=0.2\textwidth]{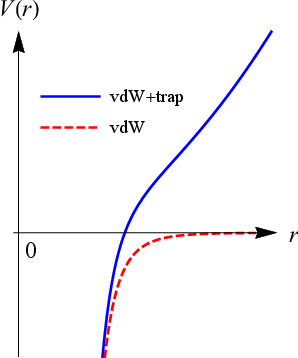}
\put(16,82){{\bf (b)}}
\end{overpic}
\caption{(color online) {\bf (a)} A schematic diagram of two atoms in a harmonic trap. {\bf (b)} Qualitative potential energy curves of two-body relative motion for the sole vdW potential (blue solid) and the vdW potential in a symmetric harmonic trap (red dashed).}
\label{fig1}
\end{figure}

For two atoms confined by identical symmetric harmonic traps, as shown in Fig.~\ref{fig1}(a), center-of-mass and relative motions can be separated. Therefore, the SE of the relative motion is
\begin{eqnarray}
\left[-\frac{\hbar^2\nabla_{\bm r}^2}{2\mu}+V(r)\right]\Psi({\bm r})=\epsilon\Psi({\bm r}),\label{se}
\end{eqnarray}
where ${\bm r}$ is the relative position and the potential [see Fig.~\ref{fig1}(b)] can be expressed as
\begin{eqnarray}
V(r)=-\frac{C_6}{r^6}+\frac{1}{2}\mu\omega^2 r^2,
\end{eqnarray}
with $\mu$ being the reduced mass of two atoms, $\epsilon$ being the energy of the relative motion, $C_6>0$ being the vdW coefficient and $\omega$ being the trapping frequency. Here, we only consider the isotropic part of the vdW potential. 

Owing to the isotropy of the potential, the angular momentum of the two-body relative motion ($\hat{L}$) and the projection to $z$-axis ($\hat{L}_z$) are conserved. The solutions of SE with $\hat{L}^2=l(l+1)\hbar^2$ and $\hat{L}_z=m\hbar$ can be expressed as $\Psi({\bm r})=\frac{u_{\epsilon l}(r)}{r}Y_l^m(\theta,\varphi)$, where $Y_l^m(\theta,\varphi)$ is spherical harmonics. The radial SE for $u_{\epsilon l}(r)$ is
\begin{eqnarray}
\left[\frac{{\rm d}^2}{{\rm d} r^2}-\frac{l(l+1)}{r^2}-\frac{r^2}{4a_h^4}+\frac{\beta_6^4}{r^6}+\bar{\epsilon}\right]u_{\epsilon l}(r)=0,
\label{rse}
\end{eqnarray}
with $a_h=\sqrt{\hbar/(2\mu\omega)}$, $\beta_6=\left(2\mu C_6/\hbar^2\right)^{1/4}$, $\bar{\epsilon}=2\mu\epsilon/\hbar^2$.

A pair of linearly independent solutions of Eq.~(\ref{rse}) can be expressed as
\begin{eqnarray}
\xi_{\epsilon l}(r)&=&\sum_{n=-\infty}^{\infty}b_n(\nu)\sqrt{r}J_{\nu+n}\left(\frac{\beta_6^2}{2r^2}\right),\label{xi}\\
\eta_{\epsilon l}(r)&=&\sum_{n=-\infty}^{\infty}(-1)^n b_n(\nu)\sqrt{r}J_{-\nu-n}\left(\frac{\beta_6^2}{2r^2}\right),\label{eta}
\end{eqnarray}
where $J_\nu(x)$ is the Bessel function of the first kind. In the following, we formulate the expression of the index $\nu$ and the coefficient $b_n(\nu)$.

\subsection{The index $\nu$}
\label{index}

The index $\nu$ is the root of the equation
\begin{eqnarray}
\det\left[{\mathcal M}(\nu)\right]=0,\label{dm}
\end{eqnarray}
where ${\mathcal M}(\nu)$ is a $2\times 2$ matrix defined as

\begin{widetext}

\begin{eqnarray}
{\mathcal M}(\nu)\equiv\frac{\Delta_h^2 \mathcal{A}^{(+)}(\nu)\mathcal{Q}^{(+)}(\nu)\mathcal{F}(\nu+2)\mathcal{A}^{(-)}(\nu+2)}{(\nu+1)(\nu+2)}-\mathcal{F}^{-1}(\nu)+\mathcal{A}(\nu)+\frac{\Delta_h^2 \mathcal{A}^{(-)}(\nu)\mathcal{Q}^{(-)}(\nu)\mathcal{F}(\nu-2)\mathcal{A}^{(+)}(\nu-2)}{\nu(\nu-1)},
\end{eqnarray}
with $\Delta_h=\beta_6^4/(256a_h^4)$. Here $\mathcal{A}^{(\pm)}(\nu)$, $\mathcal{F}(\nu)$ and $\mathcal{A}(\nu)$ are $2\times 2$ matrixed defined as
\begin{eqnarray}
\mathcal{A}^{(+)}(\nu)=\begin{pmatrix}
\displaystyle\frac{1}{\nu+1}& 0\\
-\displaystyle\frac{\Delta_\epsilon}{\Delta_h}& \displaystyle\frac{1}{\nu+3}
\end{pmatrix},&&\quad
\mathcal{A}^{(-)}(\nu)=\begin{pmatrix}
\displaystyle\frac{1}{\nu-2}& -\displaystyle\frac{\Delta_\epsilon}{\Delta_h}\\
0& \displaystyle\frac{1}{\nu}
\end{pmatrix},\\
\mathcal{F}(\nu)=\begin{pmatrix}
\displaystyle\frac{1}{\nu^2}& 0\\
0& \displaystyle\frac{1}{(\nu+1)^2}
\end{pmatrix},
&&\quad
\mathcal{A}(\nu)=\begin{pmatrix}
\displaystyle\frac{2\Delta_h}{(\nu-1)(\nu+1)}& -\displaystyle\frac{\Delta_\epsilon}{\nu+1}\\
-\displaystyle\frac{\Delta_\epsilon}{\nu}& \displaystyle\frac{2\Delta_h}{\nu(\nu+2)}
\end{pmatrix}+\nu_0^2\mathcal{I},
\end{eqnarray}
where $\nu_0=(2l+1)/4$, $\Delta_\epsilon=\bar{\epsilon}\beta_6^2/16$ and $\mathcal{I}$ is the $2\times 2$ identity matrix. The inverse of the matrix $\mathcal{F}(\nu)$ is denoted by $\mathcal{F}^{-1}(\nu)$. In addition, $\mathcal{Q}^{(\pm)}(\nu)$ are $2\times 2$ matrixes given by the continued-fraction-like recursion equations:
\begin{eqnarray}
\mathcal{Q}^{(+)}(\nu-2)&=&\bigg[\mathcal{I}-\mathcal{F}(\nu)\mathcal{A}(\nu)-\frac{\Delta_h^2 \mathcal{F}(\nu)\mathcal{A}^{(+)}(\nu)\mathcal{Q}^{(+)}(\nu)\mathcal{F}(\nu+2)\mathcal{A}^{(-)}(\nu+2)}{(\nu+1)(\nu+2)}\bigg]^{-1},\label{qv1}\\
\mathcal{Q}^{(-)}(\nu+2)&=&\bigg[\mathcal{I}-\mathcal{F}(\nu)\mathcal{A}(\nu)-\frac{\Delta_h^2 \mathcal{F}(\nu)\mathcal{A}^{(-)}(\nu)\mathcal{Q}^{(-)}(\nu)\mathcal{F}(\nu-2)\mathcal{A}^{(+)}(\nu-2)}{\nu(\nu-1)}\bigg]^{-1},\label{qv2}
\end{eqnarray}
From the above equations, $\mathcal{Q}^{(\pm)}(\nu)$ have the property
\begin{eqnarray}
\lim_{n\rightarrow\infty}\mathcal{Q}^{(\pm)}(\nu\pm n)=\mathcal{I}.
\label{qi}
\end{eqnarray}
The value of $\mathcal{Q}^{(\pm)}(\nu)$ for any $\nu$ can be evaluated by recursion based on Eqs.~(\ref{qv1}-\ref{qi}).

It should be noted that if $\nu$ is a root of Eq.~(\ref{dm}), $-\nu$, $\nu^*$ and $\nu+n$ $(n=\pm 1,\pm 2,\dots)$ are also roots of this equation. In our calculation, we choose the one that satisfies Re$(\nu)\ge 0$, Im$(\nu)\ge 0$ and $\lim_{\beta_6\rightarrow 0}\nu=\nu_0$, in contrast with Ref.~\cite{Du2022} where the last condition is related with $\lim_{\bar\epsilon\rightarrow 0}\nu$. In addition, the index $\nu$ becomes a complex number \cite{Bo1998} when $\Delta_h$ or $\Delta_\epsilon$ is beyond the respective critical values, and the real part of $\nu$ is then fixed at
\begin{eqnarray}
\mathrm{Re}(\nu)=\left\{\begin{matrix}l/2,&l\ \mathrm{even};\\(l+1)/2,&l\ \mathrm{odd}.\end{matrix}\right.
\end{eqnarray}

\subsection{The expression of $b_n(\nu)$}
To derive the expression of $b_n(\nu)$, we define the 2-component vector ${\bm B}_n(\nu)$ as
\begin{eqnarray}
{\bm B}_n(\nu)\equiv\begin{bmatrix}b_{2n}(\nu)\\b_{2n+1}(\nu)\end{bmatrix},\quad(n=0,\pm 1,\pm 2,\dots),
\end{eqnarray}
which is formulated by
\begin{eqnarray}
{\bm B}_{n}(\nu)&=&\frac{\Delta_h^n\Gamma\big(\frac{1+\nu}{2}\big)}{2^n\Gamma\big(n+\frac{1+\nu}{2}\big)}{\mathcal S}_n^{(+)}(\nu){\mathcal S}_{n-1}^{(+)}(\nu)\dots{\mathcal S}_1^{(+)}(\nu){\bm B}_0(\nu),\quad(n=1, 2, \dots),\\
{\bm B}_{-n}(\nu)&=&\frac{\Delta_h^n\Gamma\big({\rm -}\frac{\nu}{2}\big)}{(-2)^n\Gamma\big(n-\frac{\nu}{2}\big)}{\mathcal S}_n^{(-)}(\nu){\mathcal S}_{n-1}^{(-)}(\nu)\dots{\mathcal S}_1^{(-)}(\nu){\bm B}_0(\nu),\quad(n=1, 2, \dots),
\end{eqnarray}
with
\begin{eqnarray}
{\mathcal S}_n^{(\pm)}(\nu)=\mathcal{Q}^{(\pm)}(\nu\pm 2n\mp 2)\mathcal{F}(\nu\pm 2n)\mathcal{A}^{(\mp)}(\nu\pm 2n).\nonumber\\
\end{eqnarray}
Moreover, ${\bm B}_0(\nu)$, except for an overall factor, can be determined via
\begin{eqnarray}
\mathcal{M}(\nu){\bm B}_0(\nu)=0.
\end{eqnarray}

\subsection{Asymptotic behaviors} 

\subsubsection{Asymptotic behaviors for $r\rightarrow 0$}

The behaviors of solutions $\xi_{\epsilon l}(r)$ and $\eta_{\epsilon l}(r)$ in the limit $r\rightarrow 0$ can be expressed as
\begin{eqnarray}
\xi_{\epsilon l}(r\rightarrow 0)&\rightarrow&\frac{2r^{3/2}}{\sqrt{\pi}\beta_6}\left[d_{\rm c+}(\nu)\cos\left(\frac{\beta_6^2}{2r^2}-\frac{\pi}{4}\right)+d_{\rm s+}(\nu)\sin\left(\frac{\beta_6^2}{2r^2}-\frac{\pi}{4}\right)\right];\label{xi0}\\
\eta_{\epsilon l}(r\rightarrow 0)&\rightarrow&\frac{2r^{3/2}}{\sqrt{\pi}\beta_6}\left[d_{\rm c-}(\nu)\cos\left(\frac{\beta_6^2}{2r^2}-\frac{\pi}{4}\right)+d_{\rm s-}(\nu)\sin\left(\frac{\beta_6^2}{2r^2}-\frac{\pi}{4}\right)\right],\label{eta0}
\end{eqnarray}
with
\begin{eqnarray}
d_{\rm c\pm}(\nu)=\sum_{n=-\infty}^{+\infty}b_n(\nu)\cos\left[\frac{\pi(\pm\nu+n)}{2}\right];\quad d_{\rm s\pm}(\nu)=\sum_{n=-\infty}^{+\infty}b_n(\nu)\sin\left[\frac{\pi(\pm\nu+n)}{2}\right].\label{dcs}
\end{eqnarray}

\subsubsection{Asymptotic behaviors for $r\rightarrow \infty$}

The behaviors of solutions $\xi_{\epsilon l}(r)$ and $\eta_{\epsilon l}(r)$ in the limit $r\rightarrow\infty$ are given by
\begin{eqnarray}
\xi_{\epsilon l}(r\rightarrow\infty)&\rightarrow& q_-(\nu)\frac{a_h}{\sqrt{r}}\left[\frac{\Gamma(1-2\nu)}{\Gamma(\frac{1}{2}-\nu-\frac{a_h^2\bar\epsilon}{2})}\left(\frac{r^2}{2a_h^2}\right)^{-\frac{a_h^2\bar\epsilon}{2}}e^{\frac{r^2}{4a_h^2}}+\frac{\Gamma(1-2\nu)e^{-i\pi(\frac{1}{2}-\nu-\frac{a_h^2\bar\epsilon}{2})}}{\Gamma(\frac{1}{2}-\nu+\frac{a_h^2\bar\epsilon}{2})}\left(\frac{r^2}{2a_h^2}\right)^{\frac{a_h^2\bar\epsilon}{2}}e^{-\frac{r^2}{4a_h^2}}\right],\label{xii}\\
\eta_{\epsilon l}(r\rightarrow\infty)&\rightarrow& q_+(\nu)\frac{a_h}{\sqrt{r}}\left[\frac{\Gamma(1+2\nu)}{\Gamma(\frac{1}{2}+\nu-\frac{a_h^2\bar\epsilon}{2})}\left(\frac{r^2}{2a_h^2}\right)^{-\frac{a_h^2\bar\epsilon}{2}}e^{\frac{r^2}{4a_h^2}}+\frac{\Gamma(1+2\nu)e^{-i\pi(\frac{1}{2}+\nu-\frac{a_h^2\bar\epsilon}{2})}}{\Gamma(\frac{1}{2}+\nu+\frac{a_h^2\bar\epsilon}{2})}\left(\frac{r^2}{2a_h^2}\right)^{\frac{a_h^2\bar\epsilon}{2}}e^{-\frac{r^2}{4a_h^2}}\right],\label{etai}
\end{eqnarray}
where $q_\pm(\nu)$ are defined by
\begin{eqnarray}
q_\pm(\nu)\equiv q_{\pm,0}(\nu)=q_{\pm,1}(\nu),
\end{eqnarray}
and
\begin{eqnarray}
q_{\pm,\delta}(\nu)=\lim_{n\rightarrow\infty}\frac{(\pm 1)^\delta\Delta_h^{-n\mp(\nu+\delta)/2}b_{\pm 2n+\delta}(\nu)\Gamma(2n\pm\delta+1)}{2^{\pm\nu-1/2}\Gamma(\mp\nu-2n\mp\delta+1){}_2 F_1(-2n\mp\delta,1/2\pm\nu-\epsilon_h/2;1\pm2\nu;2)},\quad(\delta=0,1).
\end{eqnarray}
Here $\Gamma(z)$ is Euler's Gamma function and ${}_2 F_1(\alpha,\beta;\gamma;z)$ is the hypergeometric function. This limit of most cases is expected to be monotonically convergent based on our numerical calculation. However, the convergence speed slows down when $a_h$ becomes less than $\beta_6$, especially for high-wave and low-energy cases. We emphasize that the limit is not monotonic until the index $n$ exceeds a sufficiently large number in extreme examples.

\end{widetext}

\section{The energy spectrum} 
\label{III}

For arbitrary energy $\epsilon$ and angular momentum $l$, the solution $u_{\epsilon l}(r)$ of Eq.~(\ref{rse}) is a linear combination of solutions $\xi_{\epsilon l}(r)$ and $\eta_{\epsilon l}(r)$, which can be expressed as
\begin{eqnarray}
u_{\epsilon l}(r)=N_{\epsilon l}^{(0)}[\xi_{\epsilon l}(r)+K_{\epsilon l}^{(0)}\eta_{\epsilon l}(r)],\label{uel}
\end{eqnarray}
where $N_{\epsilon l}^{(0)}$ is a normalized factor and $K_{\epsilon l}^{(0)}$ is determined by the short-range boundary condition. Making use of Eq.~(\ref{xii}) and the long-range boundary condition of $u_{\epsilon l}(r)$, we obtain the energy spectrum, which is described by the equation
\begin{eqnarray}
K_{\epsilon l}^{(0)}=\frac{q_-(\nu)\Gamma(-2\nu)\Gamma(1/2-a_h^2\bar\epsilon/2+\nu)}{q_+(\nu)\Gamma(2\nu)\Gamma(1/2-a_h^2\bar\epsilon/2-\nu)}.\label{en}
\end{eqnarray}
As inferred from Sec. \ref{index}, the index $\nu$ is a function of $\bar\epsilon\beta_6^2$ and $a_h/\beta_6$. Therefore, the energy spectrum depends on three dimensionless parameters: $K_{\epsilon l}^{(0)}$, $\bar\epsilon\beta_6^2$ and $a_h/\beta_6$, which contain physics of the short range, the interaction and the trap.

\begin{figure}[t]
\centering
\begin{overpic}[width=0.35\textwidth]{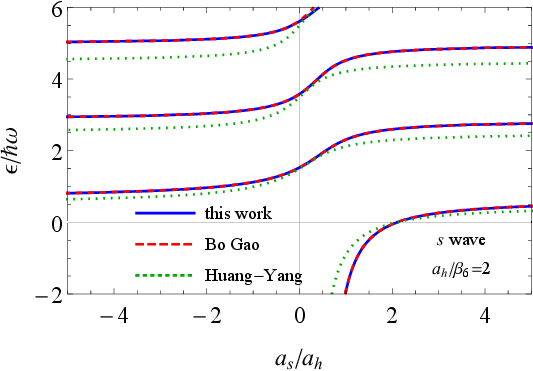}
\put(0,65){{\bf (a)}}
\end{overpic}\\
\begin{overpic}[width=0.35\textwidth]{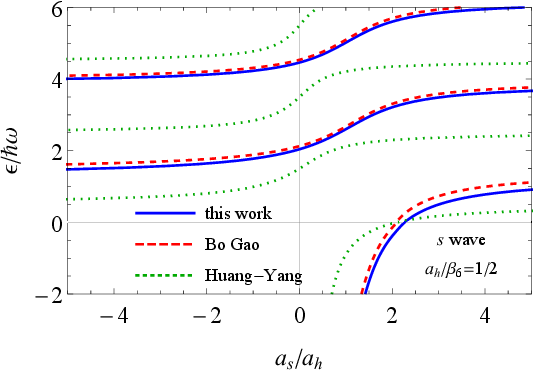}
\put(0,65){{\bf (b)}}
\end{overpic}
\caption{(color online) The $s$-wave energy spectra for {\bf (a)} $a_h/\beta_6=2$, and {\bf (b)} $a_h/\beta_6=1/2$. We show the results given by this work (blue solid), Bo Gao {\it et al.} \cite{Bo2007} (red dashed) and Huang-Yang pseudopotential (green dotted).}
\label{fig2}
\end{figure}

For cold and ultracold collisions that low-wave scattering dominates, it is more intuitive to re-express the energy spectrum in terms of $s$-wave and $p$-wave scattering length $a_s$ and $a_p$, respectively. Here we introduce $K^{(c)}$ \cite{Bo20041}, a short-range parameter insensitive to energy and angular momentum \cite{Bo2001}, as a function of $K_{\epsilon l}^{(0)}$:
\begin{eqnarray}
K^{(c)}=\frac{d_{\rm s+}(\nu)+K_{\epsilon l}^{(0)} d_{\rm s-}(\nu)}{d_{\rm c+}(\nu)+K_{\epsilon l}^{(0)} d_{\rm c-}(\nu)}.
\end{eqnarray}
$K^{(c)}$ is related to $a_s$ and $a_p$ by \cite{Bo20042}
\begin{eqnarray}
\frac{a_s}{\beta_6}&=&\frac{2\sqrt{2}\pi}{\Gamma^2(1/4)}\frac{K^{(c)}+\tan(\pi/8)}{K^{(c)}-\tan(\pi/8)},\\
\frac{a_p^3}{\beta_6^3}&=&-\frac{\Gamma(1/4)^2}{36\pi}\left[1+\frac{1+\tan(3\pi/8) K^{(c)}}{\tan(3\pi/8)-K^{(c)}}\right].
\end{eqnarray}
Thus, $K_{\epsilon l}^{(0)}$, $K^{(c)}$ and $a_{s(p)}/\beta_6$ all represent the short-range physics. In the limit of contact interaction, {\it i.e.}, $\beta_6$ is much less than other characteristic lengths, long-range physics is dominated by two parameters: $a_{s(p)}/a_h$ and $\bar\epsilon a_h^2$, which is just the main idea of pseudopotential theories.

Fig.~\ref{fig2} shows the $s$-wave energy spectra given by three theories, namely this work, Bo Gao {\it et al.} \cite{Bo2007} and Huang-Yang pseudopotential \cite{Wilkens1998} for $a_h/\beta_6=2$ and $1/2$. For $a_h/\beta_6=2$, when the length scale of interaction is comparable with the trap but is somewhat less, the spectrum derived by the pseudopotential, which actually corresponds to the limit $a_h/\beta_6\rightarrow\infty$, has significantly deviated from the other results. The difference between Bo Gao {\it et al.} and this work is roughly the order of 1\%. For $a_h/\beta_6=1/2$, when the length scale of interaction is slightly larger, the spectrum formulated by Bo Gao {\it et al.} also quantitatively fails as we have expected, especially near the threshold. Consequently, precise calculation with our method is indispensable in this case.

\begin{figure}[t]
\centering
\begin{overpic}[width=0.35\textwidth]{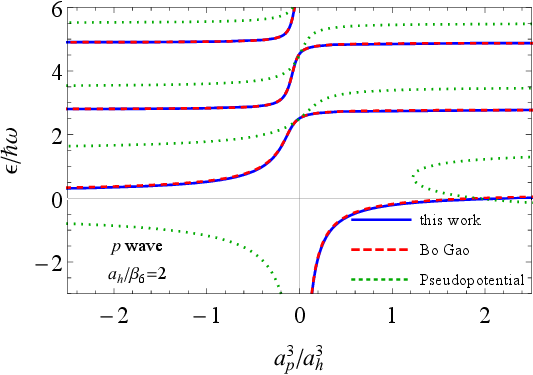}
\put(0,65){{\bf (a)}}
\end{overpic}\\
\begin{overpic}[width=0.35\textwidth]{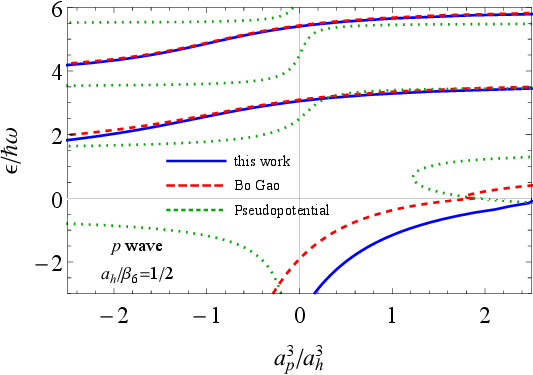}
\put(0,65){{\bf (b)}}
\end{overpic}
\caption{(color online) The $p$-wave energy spectra for {\bf (a)} $a_h/\beta_6=2$, and {\bf (b)} $a_h/\beta_6=1/2$. We also show the results given by this work (blue solid), Bo Gao {\it et al.} \cite{Bo2007} (red dashed) and pseudopotential (green dotted).}
\label{fig3}
\end{figure}

Fig.~\ref{fig3} shows the $p$-wave energy spectra for $a_h/\beta_6=2$ and $1/2$. Similarly, Bo Gao {\it et al.} is almost consistent with this work for $a_h/\beta_6=2$. In contrast, the pseudopotential fails completely to depict the spectrum while the only exception is at zero energy. This failure is a direct result from the limitation of the effective range theory. The energy-dependent scattering volume is thus introduced \cite{ZI2009, Venu2023}. In the other case with $a_h/\beta_6=1/2$, the spectrum derived by Bo Gao {\it et al.} differs more from our result compared with the $s$-wave  threshold behavior. It is remarkable that even the pseudopotential at zero energy greatly diverges from our result, so is the result of Bo Gao {\it et al.} Our theory demonstrates its importance when the length scale of the interaction is comparable with or even greater than that of the trap, especially for high-wave and low-energy cases.

\section {Discussions and summary} 
\label{IV}

This paper starts from the attractive vdW potential with $-C_6<0$, yet our method can be applied to the repulsive potential with 
$-C_6>0$. To deal with this situation, $\beta_6$ is re-defined as $\beta_6=\left(2\mu |C_6|/\hbar^2\right)^{1/4}$. Solutions of the SE $\Big[\frac{{\rm d}^2}{{\rm d} r^2}-\frac{l(l+1)}{r^2}-\frac{r^2}{4a_h^4}-\frac{\beta_6^4}{r^6}+\bar{\epsilon}\Big]u_{\epsilon l}(r)=0$ can be expressed by $\tilde\xi_{\epsilon l}(r)=\sum_{n=-\infty}^{\infty}\tilde{b}_n(\tilde\nu)\sqrt{r}I_{\tilde\nu+n}\left(\frac{\beta_6^2}{2r^2}\right)$ and $\tilde\eta_{\epsilon l}(r)=\sum_{n=-\infty}^{\infty}\tilde{b}_n(\tilde\nu)\sqrt{r}I_{-\tilde\nu-n}\left(\frac{\beta_6^2}{2r^2}\right)$. Another solution $\tilde\zeta_{\epsilon l}(r)=\sum_{n=-\infty}^{\infty}(-1)^n \tilde{b}_n(\tilde\nu)\sqrt{r}K_{\tilde\nu+n}\left(\frac{\beta_6^2}{2r^2}\right)$ just
exponentially decays in the limit $r\rightarrow 0$. The coefficient $\tilde{b}_n(\tilde\nu)$ and the index $\tilde\nu$ will be derived via the same procedure illustrated in appendixes.

Furthermore, solutions presented in this paper are also applicable to other isotropic potentials, such as $V_1(r)=-\frac{C_3}{r^3}-\frac{C_4}{r^4}$ and $V_2(r)=-\frac{C_1}{r}-\frac{C_4}{r^4}$ \cite{Fu2016}. Corresponding radial SEs can be transformed into differential equations of the same form as $V(r)$ by replacing the variable $x$ with $x_1=\sqrt{\bar\epsilon}r$ and $x_2=\sqrt{2\mu C_4/\hbar^2}/r$ respectively [Eq.~(\ref{sse})]. Thus, the above results can be used in these systems directly.

In this paper we derive a pair of special solutions $\xi_{\epsilon l}(r)$ and $\eta_{\epsilon l}(r)$ for the SE with an isotropic vdW interaction in a symmetric harmonic trap. According to asymptotic behaviors of these solutions, the energy spectrum of the two-body relative motion is obtained, which is determined by three dimensionless parameters: $K_{\epsilon l}^{(0)}$, $\bar\epsilon\beta_6^2$ and $a_h/\beta_6$. We further relate the spectrum to $s$-wave and $p$-wave scattering lengths, and hence one of the above parameters $K_{\epsilon l}^{(0)}$ is replaced by $a_{s(p)}/\beta_6$. These results are helpful to study collisions of two trapped atoms that interact with a long-range vdW potential. Moreover, solutions presented here are applicable to other two-scale isotropic potentials, which may promote the understanding of multi-scale physics.

\begin{acknowledgements}

I thank Peng Zhang for helpful discussions. This work is supported by the Fundamental Research Funds for the Central Universities, and the Research Funds of Renmin University of China No. 23XNH078.

\end{acknowledgements}

\global\long\def\id{\mathbbm{1}}
\global\long\def\ui{\mathbbm{i}}
\global\long\def\ud{\mathrm{d}}


\appendix

\begin{widetext}

\section{Derivation of solutions}
\label{solu}

In this section, we derive special solutions of Eq.~(\ref{rse}) with the method developed recently. To this end, we expand the solution as a Neumann series:
\begin{eqnarray}
u_{\epsilon l}(r)=\sqrt{r}\sum_{n=-\infty}^{+\infty}b_n(\nu)J_{\nu+n}(x),\label{neum}
\end{eqnarray}
where $x=\beta_6^2/(2r^2)$ and $J_\nu(x)$ is the Bessel function of the first kind. Substituting this form into Eq.~(\ref{rse}), we obtain
\begin{eqnarray}
\left(x^2\frac{\textrm{d}^2}{\textrm{d}x^2}+x\frac{\textrm{d}}{\textrm{d}x}+x^2-\nu_0^2+\frac{2\Delta_\epsilon}{x}-\frac{4\Delta_h}{x^2}\right)\sum_{n=-\infty}^{+\infty}b_n(\nu)J_{\nu+n}(x)=0.\label{sse}
\end{eqnarray}
with $\nu_0=(2l+1)/4$, $\Delta_\epsilon=\bar{\epsilon}\beta_6^2/16$ and $\Delta_h=\beta_6^4/(256a_h^4)$. Making use of the properties of the Bessel function:
\begin{eqnarray}
\left(x^2\frac{\textrm{d}^2}{\textrm{d}x^2}+x\frac{\textrm{d}}{\textrm{d}x}+x^2\right)J_\nu(x)=
\nu^2 J_\nu(x),\quad\frac{2}{x}J_{\nu}(x)=\frac{1}{\nu}[J_{\nu+1}(x)+J_{\nu-1}(x)],\label{bessel}
\end{eqnarray}
we further derive the recursion equation of $b_n(\nu)$ as
\begin{eqnarray}
&&[(\nu+n)^2-\nu_0^2]b_n(\nu)+\frac{\Delta_\epsilon}{\nu+n+1}b_{n+1}(\nu)+\frac{\Delta_\epsilon}{\nu+n-1}b_{n-1}(\nu)\nonumber\\
&&\quad-\Delta_h\left[\frac{b_{n-2}(\nu)}{(\nu+n-2)(\nu+n-1)}+\frac{2b_n(\nu)}{(\nu+n-1)(\nu+n+1)}+\frac{b_{n+2}(\nu)}{(\nu+n+2)(\nu+n+1)}\right]=0.\label{eqbv}
\end{eqnarray}

To formulate the expression of $b_n(\nu)$, we introduce the 2-component vector ${\bm B}_n(\nu)$, which is defined by
\begin{eqnarray}
{\bm B}_n(\nu)\equiv\begin{bmatrix}b_{2n}(\nu)\\b_{2n+1}(\nu)\end{bmatrix},\quad(n=0,\pm 1,\pm 2,\dots).
\end{eqnarray}
Thus, according to Eq.~(\ref{eqbv}), we can deduce the recursion equation of ${\bm B}_n(\nu)$:
\begin{eqnarray}
\frac{\Delta_h \mathcal{A}^{(-)}(\nu+2n){\bm B}_{n-1}(\nu)}{\nu+2n-1}-\left[\mathcal{F}^{-1}(\nu+2n)-\mathcal{A}(\nu+2n)\right]{\bm B}_n(\nu)+\frac{\Delta_h \mathcal{A}^{(+)}(\nu+2n){\bm B}_{n+1}(\nu)}{\nu+2n+2}=0,\label{eqbbv}
\end{eqnarray}
where $\mathcal{A}^{(\pm)}(\nu)$, $\mathcal{F}(\nu)$ and $\mathcal{A}(\nu)$ are $2\times 2$ matrixes defined in our main text. Then, similar to what we have done in the last paper \cite{Du2022}, we formally express ${\bm B}_n(\nu)$ as
\begin{eqnarray}
{\bm B}_{n}(\nu)&=&\frac{\Delta_h^n\Gamma\big(\frac{1+\nu}{2}\big)}{2^n\Gamma\big(n+\frac{1+\nu}{2}\big)}{\mathcal S}_n^{(+)}(\nu){\mathcal S}_{n-1}^{(+)}(\nu)\dots{\mathcal S}_1^{(+)}(\nu){\bm B}_0(\nu),\quad(n=1, 2, \dots),\\
{\bm B}_{-n}(\nu)&=&\frac{\Delta_h^n\Gamma\big({\rm -}\frac{\nu}{2}\big)}{(-2)^n\Gamma\big(n-\frac{\nu}{2}\big)}{\mathcal S}_n^{(-)}(\nu){\mathcal S}_{n-1}^{(-)}(\nu)\dots{\mathcal S}_1^{(-)}(\nu){\bm B}_0(\nu),\quad(n=1, 2, \dots),
\end{eqnarray}
with
\begin{eqnarray}
{\mathcal S}_n^{(\pm)}(\nu)=\mathcal{Q}^{(\pm)}(\nu\pm 2n\mp 2)\mathcal{F}(\nu\pm 2n)\mathcal{A}^{(\mp)}(\nu\pm 2n).
\end{eqnarray}
Substituting this form into Eq.~(\ref{eqbbv}), we obtain
\begin{eqnarray}
&&\bigg\{\left[\mathcal{Q}^{(+)}(\nu+2n-2)\mathcal{F}(\nu+2n)\right]^{-1}-\mathcal{F}^{-1}(\nu+2n)+\mathcal{A}(\nu+2n)+\Delta_h^2 \mathcal{A}^{(+)}(\nu+2n)\nonumber\\
&&\quad\times\frac{\mathcal{Q}^{(+)}(\nu+2n)\mathcal{F}(\nu+2n+2)\mathcal{A}^{(-)}(\nu+2n+2)}{(\nu+2n+1)(\nu+2n+2)}\bigg\}{\mathcal S}_n^{(+)}(\nu){\mathcal S}_{n-1}^{(+)}(\nu)\dots{\mathcal S}_1^{(+)}(\nu){\bm B}_0(\nu)=0,\quad(n=1,2,\dots);\label{eqbbv1}\\
&&\bigg\{\frac{\Delta_h^2 \mathcal{A}^{(-)}(\nu-2n)\mathcal{Q}^{(-)}(\nu-2n)\mathcal{F}(\nu-2n-2)\mathcal{A}^{(+)}(\nu-2n-2)}{(\nu-2n)(\nu-2n-1)}-\mathcal{F}^{-1}(\nu-2n)\nonumber\\
&&\quad +\mathcal{A}(\nu-2n)+\left[\mathcal{Q}^{(-)}(\nu-2n+2)\mathcal{F}(\nu-2n)\right]^{-1}\bigg\}{\mathcal S}_n^{(-)}(\nu){\mathcal S}_{n-1}^{(-)}(\nu)\dots{\mathcal S}_1^{(-)}(\nu){\bm B}_0(\nu)=0,\quad(n=1,2,\dots);\label{eqbbv2}\\
&&\bigg[\frac{\Delta_h^2 \mathcal{A}^{(+)}(\nu)\mathcal{Q}^{(+)}(\nu)\mathcal{F}(\nu+2)\mathcal{A}^{(-)}(\nu+2)}{(\nu+1)(\nu+2)}-\mathcal{F}^{-1}(\nu)+\mathcal{A}(\nu)+\frac{\Delta_h^2 \mathcal{A}^{(-)}(\nu)\mathcal{Q}^{(-)}(\nu)\mathcal{F}(\nu-2)\mathcal{A}^{(+)}(\nu-2)}{\nu(\nu-1)}\bigg]{\bm B}_0(\nu)=0.\nonumber\\\label{eqbbv0}
\end{eqnarray}
The above Eqs.~(\ref{eqbbv1}-\ref{eqbbv0}) are actually equivalent to Eq.~(\ref{eqbbv}). By taking the terms in the braces of Eqs.~(\ref{eqbbv1},\ref{eqbbv2}) to be zero, these equations can be satisfied, and hence we finally find that $\mathcal{Q}^{(\pm)}(\nu)$ are $2\times 2$ matrixes given by the continued-fraction-like recursion equation
\begin{eqnarray}
\mathcal{Q}^{(+)}(\nu-2)&=&\bigg[\mathcal{I}-\mathcal{F}(\nu)\mathcal{A}(\nu)-\frac{\Delta_h^2 \mathcal{F}(\nu)\mathcal{A}^{(+)}(\nu)\mathcal{Q}^{(+)}(\nu)\mathcal{F}(\nu+2)\mathcal{A}^{(-)}(\nu+2)}{(\nu+1)(\nu+2)}\bigg]^{-1},\\
\mathcal{Q}^{(-)}(\nu+2)&=&\bigg[\mathcal{I}-\mathcal{F}(\nu)\mathcal{A}(\nu)-\frac{\Delta_h^2 \mathcal{F}(\nu)\mathcal{A}^{(-)}(\nu)\mathcal{Q}^{(-)}(\nu)\mathcal{F}(\nu\pm2)\mathcal{A}^{(+)}(\nu-2)}{\nu(\nu-1)}\bigg]^{-1},
\end{eqnarray}
which implies
\begin{eqnarray}
\lim_{n\rightarrow+\infty}\mathcal{Q}^{(\pm)}(\nu\pm n)=\mathcal{I},
\end{eqnarray}
since $\mathcal{A}^{(\pm)}(\nu\pm 2n)$ and $\mathcal{F}(\nu\pm 2n)$ all decay for large positive $n$. As for Eq.~(\ref{eqbbv0}), a secular equation as a function of index $\nu$, the condition for a nontrival solution is that the determinant of the matrix in the braket is zero, {\it i.e.},
\begin{eqnarray}
\det\left[{\mathcal M}(\nu)\right]=0,
\end{eqnarray}
with
\begin{eqnarray}
{\mathcal M}(\nu)\equiv\frac{\Delta_h^2 \mathcal{A}^{(+)}(\nu)\mathcal{Q}^{(+)}(\nu)\mathcal{F}(\nu+2)\mathcal{A}^{(-)}(\nu+2)}{(\nu+1)(\nu+2)}-\mathcal{F}^{-1}(\nu)+\mathcal{A}(\nu)+\frac{\Delta_h^2 \mathcal{A}^{(-)}(\nu)\mathcal{Q}^{(-)}(\nu)\mathcal{F}(\nu-2)\mathcal{A}^{(+)}(\nu-2)}{\nu(\nu-1)}.\nonumber\\
\end{eqnarray}
In addition, ${\bm B}_0(\nu)$ is also resolved by Eq.~(\ref{eqbbv0}).

In summary, we have derived the index $\nu$ and the expression of $b_n(\nu)$. Therefore, Eq.~(\ref{neum}) gives a solution of Eq.~(\ref{rse}), which is just $\xi_{\epsilon l}(r)$ in our main text. Furthermore, by replacing Bessel functions of the first kind in Eq.~(\ref{neum}) with Bessel functions of the second kind, we obtain another solution $\zeta_{\epsilon l}(r)=\sum_{n=-\infty}^{+\infty}b_n(\nu)\sqrt{r}Y_{\nu+n}\left(\frac{\beta_6^2}{2r^2}\right)$. The derivation of this solution is exactly the same as above, because $Y_\nu(x)$ has identical properties as $J_\nu(x)$ in Eq.~(\ref{bessel}). Due to the relation $Y_\nu(x)=[J_\nu(x)\cos(\pi\nu)-J_{-\nu}(x)]/\sin(\pi\nu)$, we can deduce $\eta_{\epsilon l}(r)$ in our main text as a linear combination of $\xi_{\epsilon l}(r)$ and $\zeta_{\epsilon l}(r)$, and thus a solution of Eq.~(\ref{rse}).

\section{Proof of asymptotic behaviors}
\label{asym}

\subsection{Asymptotic behaviors for $r\rightarrow 0$} 

The asymptotic behavior of the Bessel function $J_{\nu}(x)$ is given by
\begin{eqnarray}
J_{\nu}(x\rightarrow\infty)\rightarrow\sqrt{\frac{2}{\pi x}}\cos\left(x-\frac{\pi\nu}{2}-\frac{\pi}{4}\right).\label{jvi}
\end{eqnarray}
The limit $r\rightarrow 0$ corresponds to $\beta_6^2/(2r^2)\rightarrow\infty$. Therefore, asymptotic behaviors of $\xi_{\epsilon l}(r)$ and $\eta_{\epsilon l}(r)$ in the limit $r\rightarrow 0$, {\it i.e.}, Eqs.~(\ref{xi0}-\ref{dcs}) in our main text, can be directly obtained by substituting Eq.~(\ref{jvi}) into Eqs.~(\ref{xi}, \ref{eta}) of our main text.

\subsection{Asymptotic behaviors for $r\rightarrow\infty$} 

The Bessel function of the first kind can be expressed as
\begin{eqnarray}
J_{\nu}(x)=\sum_{s=0}^{\infty}\frac{(-1)^s}{s!\Gamma(\nu+s+1)}\left(\frac{x}{2}\right)^{2s+\nu}.
\end{eqnarray}
Substituting this form into Eq.~(\ref{xi}), we re-express $\xi_{\epsilon l}(r)$ as
\begin{eqnarray}
\xi_{\epsilon l}(r)=\sqrt{r}\sum_{n=-\infty}^{\infty}\sum_{\delta=0,1}b_{-2n+\delta}(\nu)\sum_{s=0}^{\infty}\frac{(-1)^s \Delta_h^{-n+s+(\nu+\delta)/2}}{s!\Gamma(\nu-2n+\delta+s+1)}y^{2n-2s-\nu-\delta},\label{rexi}
\end{eqnarray}
with
\begin{eqnarray}
y\equiv\frac{r^2}{4a_h^2}.
\end{eqnarray}
Obviously, the limit $r\rightarrow\infty$ corresponds to $y\rightarrow\infty$. As before \cite{Bo1998, Bo1999, Du2022}, the right-hand-side of Eq.~(\ref{rexi}) is dominated by terms with large positive $n$ and small $s$ in the limit $y\rightarrow\infty$. Therefore, in this limit we only keep terms with positive $n$ and $s=0$, and thus obtain
\begin{eqnarray}
\xi_{\epsilon l}(r\rightarrow 0)&\rightarrow&\sqrt{r}\sum_{\delta=0,1}\sum_{n=\delta}^{\infty}b_{-2n+\delta}(\nu)\frac{\Delta_h^{-n+(\nu+\delta)/2}}{\Gamma(\nu-2n+\delta+1)}y^{2n-\nu-\delta}\nonumber\\
&\rightarrow&\frac{a_h}{\sqrt{r}}\sum_{\delta=0,1}\sum_{n=\delta}^{\infty}q_{-,\delta}(\nu)\frac{(-1)^\delta (2y)^{1/2-\nu}}{\Gamma(2n-\delta+1)}{}_2 F_1(-2n+\delta,1/2-\nu-\epsilon_h/2;1-2\nu;2)y^{2n-\delta}
\end{eqnarray}
with
\begin{eqnarray}
q_{-,\delta}(\nu)&\equiv&\lim_{n\rightarrow\infty}\frac{(-1)^\delta \Delta_h^{-n+(\nu+\delta)/2}b_{-2n+\delta}(\nu)\Gamma(2n-\delta+1)}{2^{-\nu-1/2}\Gamma(\nu-2n+\delta+1){}_2 F_1(-2n+\delta,1/2-\nu-\epsilon_h/2;1-2\nu;2)},\quad(\delta=0,1),
\end{eqnarray}
where ${}_2 F_1(\alpha,\beta;\gamma;z)$ is the hypergeometric function defined by
\begin{eqnarray}
{}_2 F_1(\alpha,\beta;\gamma;z)=\sum_{n=0}^{+\infty}\frac{(\alpha)_n(\beta)_n}{n!(\gamma)_n}z^n,
\end{eqnarray}
and $(\alpha)_n\equiv\Gamma(\alpha+n)/\Gamma(\alpha)$. It can be examined numerically that $q_{-,0}(\nu)=q_{-,1}(\nu)$, but the rigorous proof is remained to be provided. Accordingly, we define
\begin{eqnarray}
q_-(\nu)\equiv q_{-,0}(\nu)=q_{-,1}(\nu),
\end{eqnarray}
and deduce
\begin{eqnarray}
\xi_{\epsilon l}(r\rightarrow 0)&\rightarrow&q_-(\nu)\frac{a_h}{\sqrt{r}}(2y)^{1/2-\nu}\sum_{n=0}^\infty\frac{(-1)^n}{\Gamma(n+1)}{}_2 F_1(-n,1/2-\nu-\epsilon_h/2,1-2\nu,2)y^n.
\end{eqnarray}
Then, with the relation between the hypergeometric function and Whittaker $M$ function \cite{Olver2010}:
\begin{eqnarray}
M_{k,t}(z)=z^{1/2+t}\sum_{n=0}^\infty\frac{(-1)^n}{\Gamma(n+1)}{}_2 F_1\Big({\rm -}n,\frac{1}{2}+t-k;1+2t;2\Big)\left(\frac{z}{2}\right)^n
\end{eqnarray}
we obtain
\begin{eqnarray}
\xi_{\epsilon l}(r\rightarrow\infty)\rightarrow q_-(\nu)\frac{a_h}{\sqrt{r}}M_{a_h^2\bar\epsilon/2,-\nu}\left(\frac{r^2}{2a_h^2}\right).
\label{xiis}
\end{eqnarray}
The asymptotic behavior of $\eta_{\epsilon l}(r)$ is expressed as
\begin{eqnarray}
\eta_{\epsilon l}(r\rightarrow\infty)\rightarrow q_+(\nu)\frac{a_h}{\sqrt{r}}M_{a_h^2\bar\epsilon/2,\nu}\left(\frac{r^2}{2a_h^2}\right)
\label{etais}
\end{eqnarray}
with the similar deduction. The above equations are more convenient in the discussion of the energy spectrum. Finally, we give the leading term in Eqs.~(\ref{xii}, \ref{etai}) with the asymptotic behavior of Whittaker $M$ function in the limit $z\rightarrow\infty$:
\begin{eqnarray}
M_{k,t}(z\rightarrow\infty)\rightarrow\frac{\Gamma(1+2t)}{\Gamma(1/2+t-k)}z^{-k}e^{z/2}+\frac{\Gamma(1+2t)}{\Gamma(1/2+t+k)}e^{-i\pi(1/2+t-k)}z^k e^{-z/2}.
\end{eqnarray}

\section{Details for the energy spectrum} 
\label{ensp}

Making use of Eqs.~(\ref{xiis}, \ref{etais}) and the relation between Whittaker $W$ and $M$ function: 
\begin{eqnarray}
W_{k,t}(z)=\frac{\Gamma (-2 t)}{\Gamma (1/2-k-t)}M_{k,t}(z)+\frac{\Gamma (2 t)}{\Gamma (1/2-k+t)}M_{k,-t}(z),
\end{eqnarray}
we can derive the large-$r$ asymptotic behavior of $u_{\epsilon l}(r)$ in Eq.~(\ref{uel}) as
\begin{eqnarray}
u_{\epsilon l}(r\rightarrow\infty)&\rightarrow& N_{\epsilon l}^{(0)}\frac{a_h}{\sqrt{r}}\bigg\{q_-(\nu)\frac{\Gamma(1/2-a_h^2\bar\epsilon/2+\nu)}{\Gamma(2\nu)}W_{a_h^2\bar\epsilon/2,\nu}\left(\frac{r^2}{2a_h^2}\right)\nonumber\\
&&+\bigg[K_{\epsilon l}^{(0)} q_+(\nu)-q_-(\nu)\frac{\Gamma(-2\nu)\Gamma(1/2-a_h^2\bar\epsilon/2+\nu)}{\Gamma(2\nu)\Gamma(1/2-a_h^2\bar\epsilon/2-\nu)}\bigg]M_{a_h^2\bar\epsilon/2,\nu}\left(\frac{r^2}{2a_h^2}\right)\bigg\}.
\end{eqnarray}

\end{widetext}
The asymptotic behavior of $W_{k,t}(z)$ in the limit $z\rightarrow\infty$ is
\begin{eqnarray}
W_{k,t}(z\rightarrow\infty)\rightarrow z^k e^{-z/2}.
\end{eqnarray}
Thus, $W_{k,t}(z)$ behaves approximately as an exponential decay function in the limit $z\rightarrow\infty$ while $M_{k,t}(z)$ contains both increasing and decay parts. To satisfy the long-range boundary condition $u_{\epsilon l}(r\rightarrow\infty)\rightarrow 0$ for the two-body relative motion in a trap, we further deduce the energy spectrum as
\begin{eqnarray}
K_{\epsilon l}^{(0)}=\frac{q_-(\nu)\Gamma(-2\nu)\Gamma(1/2-a_h^2\bar\epsilon/2+\nu)}{q_+(\nu)\Gamma(2\nu)\Gamma(1/2-a_h^2\bar\epsilon/2-\nu)}.\label{sen}
\end{eqnarray}
The same results can be obtained via the leading terms of asymptotic behaviors in Eqs.~(\ref{xii}, \ref{etai}).

In order to express the energy spectrum with scattering lengths, we introduce another pair of linearly independent solutions $\xi_{\epsilon l}^{(c)}(r)$ and $\eta_{\epsilon l}^{(c)}(r)$, which are defined by asymptotic behaviors \cite{Bo20041}
\begin{eqnarray}
\xi_{\epsilon l}^{(c)}(r\rightarrow 0)&\rightarrow&\frac{2r^{3/2}}{\sqrt{\pi}\beta_6}\cos\left(\frac{\beta_6^2}{2r^2}-\frac{\pi}{4}\right),\label{xic0}\\\eta_{\epsilon l}^{(c)}(r\rightarrow 0)&\rightarrow&\frac{2r^{3/2}}{\sqrt{\pi}\beta_6}\sin\left(\frac{\beta_6^2}{2r^2}-\frac{\pi}{4}\right).\label{etac0}
\end{eqnarray}
According to this pair of solutions, $u_{\epsilon l}(r)$ can be re-expressed as
\begin{eqnarray}
u_{\epsilon l}(r)=N_{\epsilon l}^{(c)}[\xi_{\epsilon l}^{(c)}(r)+K^{(c)}\eta_{\epsilon l}^{(c)}(r)].
\end{eqnarray}
Combining Eqs.~(\ref{xi0},\ref{eta0}) and Eq.~(\ref{xic0}, \ref{etac0}), we obtain the relation between $K^{(c)}$ and $K_{\epsilon l}^{(0)}$:
\begin{eqnarray}
K^{(c)}=\frac{d_{\rm s+}(\nu)+K_{\epsilon l}^{(0)} d_{\rm s-}(\nu)}{d_{\rm c+}(\nu)+K_{\epsilon l}^{(0)} d_{\rm c-}(\nu)}.
\end{eqnarray}

\bibliography{trapvdw}

\end{document}